\documentclass[aps,twocolumn,superscriptaddress,showpacs,amsmath]{revtex4}
\usepackage{graphicx}
\usepackage{bm}
\usepackage{amsmath}
\begin{document}

\title{Entanglement and Quantum Phase Transition in Low Dimensional Spin Systems}

\author{Yan Chen$^1$, Paolo Zanardi$^2$, Z. D. Wang$^{1,3}$, and F. C. Zhang}

\affiliation{Department of Physics and Center of Theoretical and
Computational Physics, University of Hong Kong, Pokfulam Road,
Hong Kong, China\\$^2$Institute for Scientific Interchange
Foundation, Viale Settimio Severo 65, I-10133 Torino, Italy\\
$^3$National Laboratory of Solid State Microstructures, Nanjing
University, Nanjing, China}

\date{\today}

\begin{abstract}
Entanglement of the ground states in $XXZ$ and dimerized
Heisenberg spin chains as well as in a two-leg spin ladder is
analyzed by using the spin-spin concurrence and the entanglement
entropy between a selected sublattice of spins and the rest of the
system. In particular, we reveal that quantum phase transition
points/boundaries may be identified  based on the analysis on the
local extreme of this entanglement entropy, which is illustrated
to be superior over the concurrence scenario and may enable us to
explore quantum phase transitions in many other systems including
higher dimensional ones.

\end{abstract}


\maketitle

{\em Introduction}. Quantum entanglement has attracted a lot of
attention for its potential applications in  quantum information
processing~\cite{Wootters98,Nielsen1}. More recently,  entanglement
has also been recognized to play an important role in the study of
quantum many particle physics, and experimental measurements have
demonstrated that  it can  affect the macroscopic properties of
solids~\cite{Aeppli}.
There have been a number of theoretical studies of the entanglement
and quantum phase transitions~\cite{Sachdev00} in one dimensional
spin systems and in interacting fermion and boson
systems~\cite{Connor01,Zanardi02,Wang02,Santos03,Gu03,Cirac04,Pachos04,Martin04,Wei04,Yang04,Gu04,Zanardi03}.
These studies showed that the entanglement can exhibit a non-trivial
behavior in condensed matter systems, such as the coincidence of the
singularity in the entanglement and quantum phase transition point
in certain systems~\cite{AOsterloh2002}. Most of the previous works
were mainly focused on spin chains and the spin-spin concurrence
\cite{Wootters98} was used to describe (two-particle) the
entanglement.  The scaling behavior of entanglement between a block
of contiguous spins and the rest of the system 
in a spin chain has been studied both near and at the
quantum critical point~\cite{GVidal03}, and a connection with
quantum phase transitions was elaborated.

In this paper, 
we adopt a preferable and distinct way to partition the system  to
study the ground state of three low dimensional spin systems: the
XXZ spin chain, dimerized Heisenberg chain, and two-leg spin
ladder. We find that the entanglement entropy is superior over
concurrence in revealing quantum transition points. In the
dimerized Heisenberg system, this entanglement entropy has maxima
and minima which are in one-to-one correspondence with the
transition points, while the concurrence fails to locate them.
Even though
the spin ladder system exhibits a complex  phase structure, 
 our entanglement entropy scenario is likely to enable
us to identify the phase boundaries. It is expected that our
approach is more efficient and powerful in exploring quantum phase
transitions in various systems.

{\em The models}. We here study the entanglement in three types of
spin-$1/2$ systems: XXZ chain, dimerized Heisenberg chain, and
two-leg XXZ ladder, with the Hamiltonian given by
\begin{eqnarray}
H_{XXZ} &=& J \sum_{i=1}^{N} \left(\mbox{\boldmath$S$}_i^x
\mbox{\boldmath$S$}_{i+1}^x + \mbox{\boldmath$S$}_i^y
\mbox{\boldmath$S$}_{i+1}^y + \Delta \mbox{\boldmath$S$}_i^z
\mbox{\boldmath$S$}_{i+1}^z \right),\\
H_{D} &=& \sum_{i=1}^{N/2} \left(J_1 \mbox{\boldmath$S$}_{2i-1}
\cdot \mbox{\boldmath$S$}_{2i} + J_2 \mbox{\boldmath$S$}_{2i}
\cdot \mbox{\boldmath$S$}_{2i+1} \right), \\
H_{L} &=& J \sum_{i=1}^2\sum_{j=1}^N
\left(\mbox{\boldmath$S$}_{i,j}^x \mbox{\boldmath$S$}_{i,j+1}^x +
\mbox{\boldmath$S$}_{i,j}^y \mbox{\boldmath$S$}_{i,j+1}^y + \Delta
\mbox{\boldmath$S$}_{i,j}^z \mbox{\boldmath$S$}_{i,j+1}^z
\right)\nonumber \\
&&  +J' \sum_{j=1}^{N} \left(\mbox{\boldmath$S$}_{1,j}^x
\mbox{\boldmath$S$}_{2,j}^x + \mbox{\boldmath$S$}_{1,j}^y
\mbox{\boldmath$S$}_{2,j}^y + \Delta \mbox{\boldmath$S$}_{1,j}^z
\mbox{\boldmath$S$}_{2,j}^z \right),
\end{eqnarray}
where $\mbox{\boldmath$S$}_{i}$ is the spin operator, $N$ is the
linear dimension of the lattice (we adopt periodic boundary
conditions). The anisotropy parameter is denoted by $\Delta$ in
the $XXZ$ model. $J_1$ and $J_2$ are two nearest-neighbor exchange
couplings in the dimerized Heisenberg Hamiltonian. In the ladder
system, the Hamiltonian contains intersite interactions along the
chains ($J$) and  the rungs ($J'$).

To quantify the entanglement in these spin systems, we first adopt
concurrence to describe two-spin entanglement \cite{Wootters98}.
This quantity can be calculated from correlation functions
$G^{\alpha\alpha}:=<\sigma^\alpha\otimes\sigma^\alpha>,\,(\alpha=x,y,z)$
\cite{Zanardi02}:
\begin{eqnarray}
C=\frac{1}{2}\max\Bigl[\,0,2|G^{xx}+G^{yy}|-G^{zz}-1\,\Bigr],
\end{eqnarray}
where the two site indexes are omitted. Since the correlation
function decays rapidly, the concurrence is nonzero only between
two closer sites. Therefore, the concurrence seems to provide
limited information on the quantum phase transition. In a
different approach,  Vidal {\it et al}.~\cite{GVidal03}
studied the  entanglement between
a block of $L$ contiguous spins $B_L$ and the rest of the system
$R_L$. The entropy of entanglement for the ground state $\Psi_g$
is  given by,
\begin{eqnarray}
S_L := - \rm{tr} \left( \rho_{L} \log_2 \rho_{L} \right),
\label{ent-ent}
\end{eqnarray}
where $\rho_L \equiv \rm{tr}_{B_L}
|{\Psi_g}\rangle\langle{\Psi_g}|$ is the reduced density matrix
for $B_L$, a block of $L$ spins. The scaling of the entanglement
in the block size $L$ may allow one to establish a  connection
between $S_L$ and the quantum phase transition: $S_L\sim \ln L$ at
the critical point, and $S_L\sim \rm{const}$ away from the
critical point; while this idea is applicable to few one
dimensional models. Here we adopt a distinct way that
bi-partitions the system into two sublattices $B_L$ and $R_L$,
with the largest number of connecting bonds between
them~\cite{Nature,footnote0}. The motivation of such partition is
to best reveal the correlation between the two sublattices. We
will see that in most cases the considered entanglement entropy
between the two subsystems  scales linearly with the size of the
{\em boundary} between them [notice that in the case
\cite{GVidal03}, the boundary is zero-dimensional i.e., one single
spin]\cite{geom-ent}.

All the above three Hamiltonians have a rotational symmetry around
the z-axes. The calculations presented below are carried out in an
invariant subspace with $S_z=0$. Lanczos algorithms are employed
to calculate the ground state $|\Psi_g\rangle$, from which we
construct the density matrix for the whole system. We then obtain
the reduced density matrix $\rho_L$ by tracing out $R_L,$ and
compute its Von Neumann entropy in Eq.(5).

\begin{figure}[t]
\includegraphics[width=6.1cm]{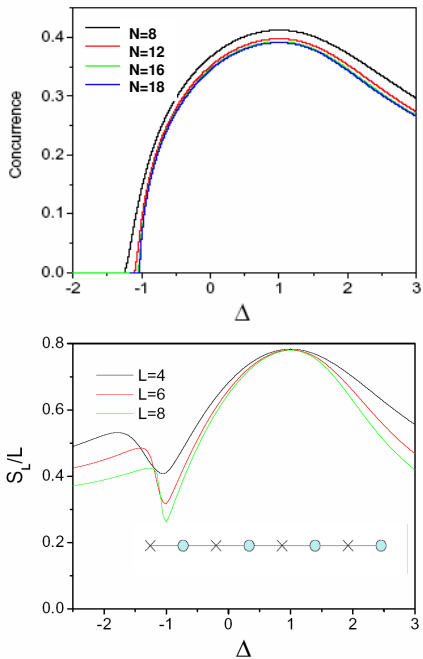}
\caption{\label{Fig1} The ground state concurrence (a) and the
entanglement entropy per site $S_L/L$ (b) in terms of $\Delta$ in
XXZ spin chain for various lattice sizes $(N=2L)$. The inset of
(b) displays the sublattice bipartition.}
\end{figure}
{\em $XXZ$-chain}. Let us first consider the one-dimensional XXZ
model. As a simple toy model, a great deal of work has been
devoted to analyze its entanglement and quantum phase
transition~\cite{book1}. It is well known that $\Delta=1$ is the
antiferromagnetic isotropic point, $\Delta=-1$ is the
ferromagnetic isotropic point, and $\Delta=0$ is the pure XY
point. This allows us to describe the various domains as a
function of $\Delta$. The system is in an Ising ferromagnetic
phase at $\Delta<-1$, Ising antiferromagnetic phase at $\Delta>1$,
and XY phase at $-1< \Delta<1$.  In Fig. 1(a), we show the
pairwise concurrence as a function of anisotropy $\Delta$ for
different number of sites $L$. Because the concurrence here is
expressed in terms of the two-site correlation function of the
nearest-neighbor sites, its value converges quickly as $L$
increases. As we can see, the concurrence at one phase transition
point $\Delta=1$ reaches the maximum while the concurrence emerges
at another transition point $\Delta=-1$ as $L$ approaches
infinity. These two quantum phase transition points can be
identified from the analysis of concurrence in this case. We now
address  the entanglement entropy of the sublattice $B_L=\{
\rm{odd-sites}\}$ with $L=N/2$~\cite{footnote}. Fig. 1(b) displays
the numerical results of the entanglement entropy $S_L/L$ as a
function of $\Delta$.
We obtain the reduced subsystem of crosses by tracing out the spin
degree of freedom on circle points. As the system is in the
vicinity of the quantum phase transition point, we may expect
$S_L/L$ to reach its extreme value. We find that the transition
points $\Delta=1$ and $\Delta=-1$ correspond to the maximum and
minimum of the sublattice entanglement entropy.
As for the latter case, a simple analysis of the scaling behavior
around the minimum point shows that the location of the transition
point approaches to $\Delta=-1$ and $S_L/L=0$ as the size of the
subsystem increases. Thus at $L \rightarrow \infty$, $S_L/L
\rightarrow 0$ for $\Delta<-1$. Therefore, the two transition points
can be clearly specified and a distinct connection  between quantum
phase transition  and the entanglement entropy   has been
established.

\begin{figure}[t]
\includegraphics[width=6.1cm]{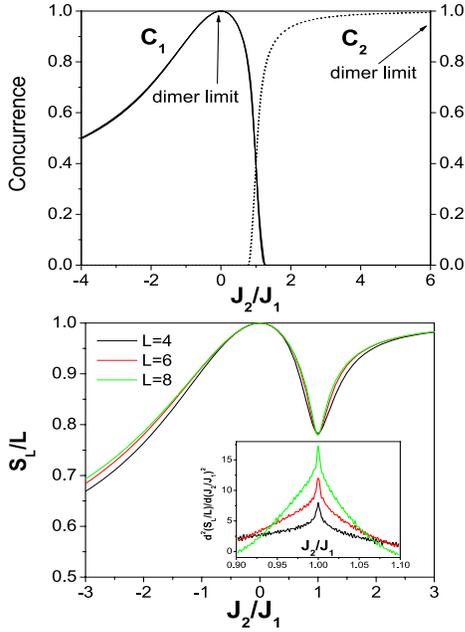}
\caption{\label{Fig2} For various lattice sizes $N$ in a dimerized
Heisenberg chain, (a) the ground state concurrences $C_1$ (solid
line) and $C_2$ (dotted line) versus $\Delta$, and (b) $S_L/L$ as
a function of the sublattice size $L$. The inset in (b) shows the
divergence of the $d^{2}(S_L/L)/d(J_2/J_1)^{2}$ at the block
entanglement minimum point as the system size increases.}
\end{figure}
{\em Dimerized Heisenberg chain}. We now consider the dimerized
Heisenberg chain~\cite{MHase93}; this model is characterized by an
alternation of strong and weak bonds between two nearest neighboring
spins.
In the case of $0<J_2/J_1\ll 1$, the ground state is just an
ensemble of $N/2$ uncoupled dimers around the strong bonds.
Consequently, there is an energy gap of order of $J_1$ to separate
the singlet ground state ($S_z=0$) from the first excited state with
$S_z=1$. All the spins are locked into singlet states. At
$J_2/J_1=1$, the system is reduced to the isotropic
antiferromagnetic Heisenberg chain that is quasi-long-range ordered,
and belongs to a different universality class from the dimerized
system. In the case of $J_2/J_1<0$, the ground state is a product of
spin singlet pairs coupled by ferromagnetic bonds. We can define two
concurrences $C_1$ and $C_2$, which correspond to two
nearest-neighboring spins coupled by bonds $J_1$ and $J_2$,
respectively. The pairwise concurrence as a function of $J_2/J_1$
for different lattice size $L$ is shown in Fig. 2(a). It can be
easily seen that the value of concurrence is independent of the
lattice size. In the limit of strong dimerization, $J_2/J_1=0$ (or
equivalently $J_2/J_1 \rightarrow \infty$), the concurrence $C_1$
($C_2$) reaches a maximum value. However, the concurrence analysis
does not enable us to identify the transition point $J_2/J_1=1$
unambiguously. Below, we examine the entanglement entropy of the
chosen  odd-site sublattice (or the enen-site sublattice). In this
case,  the two transition points are found to correspond to a local
maximum and a local minimum in the entanglement entropy [see Fig.
2(b)].
Here it is interesting to  notice that the second derivative of
$S_L/L$ with respect to $J_2/J_1$ seems to exhibit singularity at
the minimum point $J_2/J_1=1$ (see Fig. 2(b) inset), which
corresponds to the only gapless point. Due to the limitation of
our computational resources, a detailed scaling analysis is not
available.
In addition, it is seen that $S_L/L$ is weakly size dependent and
 converges quickly to a fixed value at the transition point with
increasing lattice size.

\begin{figure}[t]
\vspace{0.20cm}
\includegraphics[width=6.5cm]{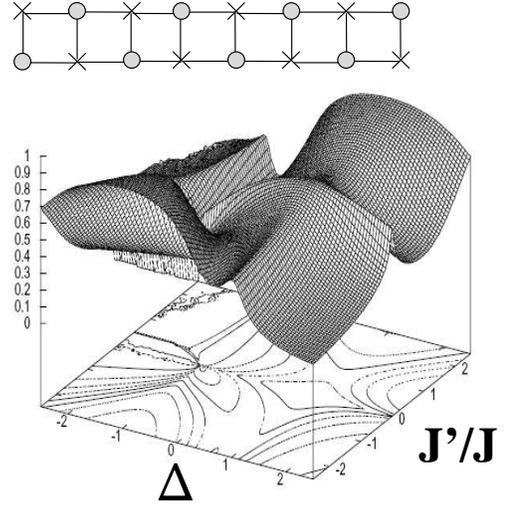}
\caption{\label{Fig3} $S_L/L$ $(L=8)$ as a function of the $J'/J$
and $\Delta$ in a coupled 2-leg XXZ spin ladder system.  The results
for $L=6$ are essentially same.}
\end{figure}
{\em Spin ladders}. Spin ladders have been attracting great interest
in the field of low dimensional spin systems in recent several years
~\cite{Rice95}. The physics of Heisenberg ladders can  be understood
in some special limits. For large and negative $J'$, the 2-leg
ladder is equivalent to an $S=1$ spin chain, and the system is
gapful~\cite{Haldane88}. For large and positive $J'$, the two spins
along the rung are locked into a singlet state. The ground state is
a collection of spin singlets on each rung and is gapful. The phase
diagram of two coupled XXZ chains was previously determined by
utilizing Hamiltonian mappings and Abelian
bosonization~\cite{Strong92}. Previous studies presumably indicated
five distinct phases in this complicated phase diagram as a function
of $\Delta$ and $J'/J$. We first examine the dependence on the
coupling constant  of two nearest-neighbor spin concurrence: this
analysis failed to describe the rich structure of the  phase
diagram. Below we report our study on  the  entanglement entropy of
the chosen subsystem and its relevance to quantum phase transition
point.  The spatial profiles and contour plots of $S_L/L$ as a
function of $J'/J$ and $\Delta$ are displayed in Fig. 3. From our
analysis from above-mentioned two models, we know that both the
ridges and valleys may correspond to possible phase boundaries. For
example, one can distinguish the boundary $J'/J=0$ (ridge or
valley), $\Delta=1$ for $J'/J<0$ (ridge), $\Delta=-1$ for $J'<0$
(ridge).
Derived from the numerical results, Fig. 4 shows the schematic phase
boundaries of a coupled 2-leg XXZ spin ladder system (the dashed
line denotes the ridge of its derivative). Most phase boundaries are
in agreement with those in Ref. 26. Our results suggest the
existence of a new quantum phase in the region $\Delta<-1.5$ and
$J'/J>0.5$, which has not been reported in previous
studies~\cite{Strong92,Orignac98,footnote1}.

Finally, it is important to emphasize that, in comparison with the
previous investigations~\cite{GVidal03},
we do not have to deal with a large system i.e., a large $L$, to
capture its scaling behavior. This indicates that the finite size
effects are not important in our approach; numerical results for a
quite small system may disclose relevant information about the
quantum phase transition points. In this sense, the present
approach is broadly applicable to many other systems including
spin and electron systems in higher dimension~\cite{Nature}, not
limited merely to the spin-chains.

\begin{figure}[t]
\includegraphics[width=5.5cm]{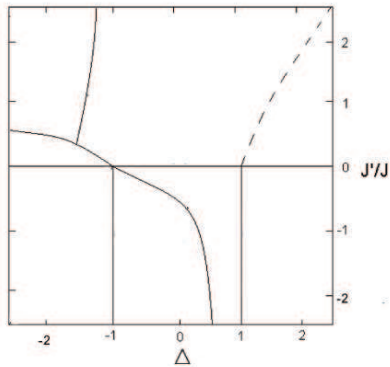}
\caption{\label{Fig4} Schematic phase boundaries of a coupled 2-leg
XXZ spin ladder system.}
\end{figure}
{\em Summary}. We have investigated the ground-state entanglement in
three one-dimensional spin systems by using both two-spin
concurrence and entanglement entropy of the preferably chosen
sublattice. Concurrence may provide some partial insights about
certain quantum phase transition points. However, our
 entanglement entropy scenario allows us to establish a
distinct connection between its local maxima/minima and transition
points, which is promising for
 shedding a new light on the
understanding of quantum phase transition and quantum
entanglement.

We are grateful to Y. Q. Li, T. K. Lee, F. Mila and X. Q. Wang for
useful discussions. This work was supported by the
RGC grants of Hong Kong, the HKU URC Seed Funding, and the NSFC
under Grant No. 10429401.


\begin{thebibliography}{}

\bibitem{Wootters98} W.K. Wootters, Phys. Rev. Lett. {\bf 80}, 2245 (1998).

\bibitem{Nielsen1} M.A. Nielsen and I.L. Chuang, {\it Quantum Computation and
Quantum Information} (Cambridge University Press, Cambridge,
England, 2000).

\bibitem{Aeppli} S. Ghosh, {\em et al.}, Nature {\bf 425}, 48 (2003).

\bibitem{Sachdev00} S. Sachdev, {\it Quantum Phase Transitions} (Cambridge University
Press, Cambridge, England, 2000).

\bibitem{Connor01}  K.M. O'Connor and W.K. Wootters, Phys. Rev. A {\bf 63}, 052302
(2001).

\bibitem{Zanardi02} P. Zanardi, Phys. Rev. A {\bf 65}, 042101(2002).

\bibitem{Wang02} X. Wang, and P. Zanardi, Phys. Lett. A {\bf 301}, 1(2002).

\bibitem{Santos03} L.F. Santos, Phys. Rev. A {\bf 67}, 062306 (2003).

\bibitem{Gu03} S.J. Gu, H.Q. Lin, and Y.Q. Li, Phys. Rev. A {\bf 68}, 042330
(2003).

\bibitem{Cirac04} F. Verstraete, M.A. Martin-Delgado, and J.I.
Cirac, Phys. Rev. Lett. {\bf 92}, 087201 (2004).

\bibitem{Pachos04} J.K. Pachos and M.B. Plenio, Phys. Rev. Lett. {\bf 93}, 056402
(2004).

\bibitem{Martin04} J.J. Garcia-Ripoll, M.A. Martin-Delgado, and J.I.
Cirac, quant-ph/0404566.

\bibitem{Wei04} T.C. Wei, {\em et al.}, quant-ph/0405162.

\bibitem{Yang04} M.F. Yang, Phys. Rev. A {\bf 71}, 030302 (2005).

\bibitem{Gu04} S.J. Gu, S.S. Deng, Y.Q. Li, and H.Q. Lin, Phys. Rev. Lett. {\bf 93}, 086402
(2004).

\bibitem{Zanardi03} P. Giorda, and P. Zanardi, quant-ph/0311058.


\bibitem{AOsterloh2002} A. Osterloh, {\em et al.}, Nature {\bf
416}, 608 (2002).


\bibitem{GVidal03}
G. Vidal, J.I. Latorre, E. Rico, and A. Kitaev, Phys. Rev. Lett,
{\bf 90}, 227902 (2003).

\bibitem{Nature} Y. Chen, Z.D. Wang, F.C. Zhang (unpublished).

\bibitem{footnote0} This entanglement entropy satisfies the subadditivity property, see for example,
A. Wehrl, Rev. Mod. Phys. {\bf 50}, 221 (1978).



\bibitem{geom-ent} M.B. Plenio, {\em et al.}, quant-ph/0405142; A. Hamma, R. Ionicioiu, and
P. Zanardi, quant-ph/0406202.


\bibitem{book1} T. Giamarchi, {\it Quantum Physics in One Dimension} (Oxford
Science Publications, Oxford, 2004)

\bibitem{footnote} We have examined the entanglement between
the subsystem of $L$ contiguous spins and its environment, and
found that such a partition does not provide clear
connection with the quantum critical point.

\bibitem{MHase93} M. Hase, I. Terasaki, and K. Uchinokura, Phys. Rev. Lett. {\bf
70}, 3651 (1993).




\bibitem{Rice95} E. Dagotto and T.M. Rice, Science, {\bf 271}, 618 (1996).

\bibitem{Haldane88} F.D.M. Haldane, Phys. Rev. Lett. {\bf 60}, 635 (1988).

\bibitem{Strong92} S.P. Strong and A.J. Millis, Phys. Rev. Lett. {\bf 69}, 2419
(1992).

\bibitem{Orignac98} E. Orignac and T. Giamarchi, Phys. Rev. B, {\bf 57}, 5812 (1998).

\bibitem{footnote1} It is quite clear each chain is in Ising ferromagnetic phase at
$\Delta \ll -1$. For large and positive $J'$, the two spins along
the rung are locked into singlet state. So this new phase may have
two ferromagnetic chains with spin singlet pair along the rung
direction.

\end{thebibliography}
\end{document}